\documentclass[aps,pra,groupedaddress,preprint]{revtex4}
\usepackage{latexsym,amsfonts}
\usepackage{hyperref}

\begin{document}     
\title{On the integral law of thermal radiation\footnote{\href{http://dx.doi.org/10.1134/S1061920814040049}{Russian Journal of Mathematical Physics, {\bf 21} (4), 460-471 (2014),}\\ \href{http://hdl.handle.net/11858/00-001M-0000-0024-756D-8}{hdl.handle.net/11858/00-001M-0000-0024-756D-8}, preprint IHES P-14-43\href{http://preprints.ihes.fr}{preprints.ihes.fr}}}
\author{Yuri V. Gusev}
\affiliation{Lebedev Research Center in Physics, Leninskii Prospekt 53, Moscow 119991, Russia}
\affiliation{Max Planck Institute for Gravitational Physics (Albert Einstein Institute), Am M\"uhlenberg 1, D-14476 Golm, Germany}
\affiliation{Institut des Hautes \'Etudes Scientifiques, Le Bois-Marie, 35 route de Chartres, 91440 Bures-sur-Yvette, France}
\date{May 25, 2014}

\begin{abstract}
The integral law of thermal radiation by finite size emitters is studied. Two geometrical characteristics of a radiating body or a cavity, its volume and its boundary area, define two terms in its radiance. The term defined by the volume corresponds to the Stefan-Boltzmann law. The term defined by the boundary area is proportional to the third power of temperature and inversely proportional to the emitter's effective size, which is defined as the ratio of its volume to its boundary area. It is shown that the cubic temperature contribution is observed in experiments. This term explains the intrinsic uncertainty of the NPL experiment on radiometric determination of the Stefan-Boltzmann constant. It is also quantitatively confirmed by data from the NIST calibration of cryogenic blackbodies. Its relevance to the size of source effect in optical radiometry is proposed. The conjecture that this generalized law is valid for arbitrary temperature and effective size is supported by the experiments on thermal emission from nano-heaters.
\end{abstract}

\maketitle


It is well known that the laws of thermal radiation, the laws of Planck, Stefan-Boltzmann and Wien, in their classic form, are not exact in describing thermal phenomena. These deviations appear even under ideal physical conditions designed in metrology labs, in contrast to realistic physical systems, e.g. \cite{Siegel-book1992}. They always exist because any radiating body or cavity (emitter) has a finite size and a shape. The finite size effects are usually neglected in the textbook derivations of thermal radiation laws \cite{Sommerfeld-bookV5,Landau-V5}, and their magnitude is assumed to be well below experimental precisions \cite{BaltesHilf-book1976,Garcia-PRA2008}. However, in this paper we show that a deviation, several orders of magnitude larger at present metrological conditions, first conjectured by H. Weyl \cite{Weyl-JRAM1913} in his theory of thermal radiation  \cite{Weyl-MA1912} and derived by V.P. Maslov in new statistical physics  \cite{Maslov-TMF2008,Maslov-MN2008}, is measured in experiments. Our theoretical proposal is motivated by the field theory and it also follows from a simple dimensional analysis. The discussed formula is in agreement with a number of experiments and observations. It could be applicable to the whole domain of thermal radiation phenomena.

\section{Thermal radiation of finite size emitters} \label{thermal}

When experiments are performed at the nanowatt and nanometer scale, and the blackbody antennas \cite{Schuller-NPhot2009} and the coherent thermal radiation sources \cite{Greffet-Nature2002} are offered, it is important to understand fundamental limitations on the laws of thermal radiation, as they may not be adequate in their classic form.  Under typical conditions in metrology, the finite size modifications are negligible (equal to or smaller than the measurement uncertainty), and the Stefan-Boltzmann, Planck and Wien laws are accepted in their ideal form. This is not the case for arbitrary physical systems. Therefore, in the last two decades it has become obvious that these laws often cannot describe experimental observations. To mitigate these discrepancies, metrologists introduced empirical corrections. The examples are the generalized Wien's law \cite{Fisenko-JPDA1999} and the modified Planck's law \cite{Fischer-BIPM2008}. The  Bureau International des Poids et Mesures (International Bureau of Weights and Measures) recognized that the Planck's law is not accurate at temperatures below the silver point (infrared radiation) and recommended using the Sakuma-Hattori equations \cite{Fischer-BIPM2008,Sounders-Metro2003} instead of the Planck's equation. These changes should lead naturally to modifications of the integral law of thermal radiation, the Stefan-Boltzmannn law, but apparently no such empirical correction was suggested, even though the deviations are significant \cite{Carter-Metro2006} or even larger than the Stefan-Boltzmann radiation \cite{Ingvarsson-OC2007}. Below we show that observed violations of the Stefan-Boltzmann law are caused by the finite size modification, which was customarily not taken into account. 

In radiometry, a radiation detector is  measuring the radiant power $\Phi$ through its aperture. From these observations, the source's radiance, the radiating power per the emitter's unit area per steradian, is deduced.  In the integral law of thermal radiation law, known as the Stefan-Boltzmann law \cite{Sommerfeld-bookV5,Landau-V5}, enters the {\em radiant exitance} $j$ (in $W/m^2$), which is found from the radiant power $\Phi$ divided by the dimensionless configuration factor $B$, (determined from the distance between apertures and the area of a detector's aperture) and by the area of the blackbody's aperture $A$,
\begin{equation}
j = \Phi/(B A). \label{jPhi}
\end{equation} 
The radiant exitance is the radiant power per unit area of the source emitted into a hemisphere around the source, assuming a Lambertian source that emits equally in all directions \cite{Parr-book2005}. For more explanations and details of the current theory and experiments in optical radiometry, see the book \cite{Parr-book2005}. 

The Stefan-Boltzmann law in its textbook form states that the radiant exitance is \cite{Sommerfeld-bookV5,Landau-V5},
\begin{equation}
j_{\mathsf{SB}} = \sigma_{\mathrm{th}} T_{\mathsf{SB}}^4, \label{Tsb}
\end{equation}
where  the Stefan-Boltzmann constant $\sigma_{\mathrm{th}}$ defined as,
\begin{equation}
\sigma_{\mathrm{th}} \equiv \frac{\pi^2}{60} \frac{{k_{\mathsf{B}}}^4}{\hbar^3 c^2}. \label{sigma}
\end{equation}
The currently recommended by  CODATA  value  of the Stefan-Boltzmann constant  \cite{CODATA-RMP2012} is not defined by direct measurements but expressed through the molar gas constant (\ref{sigma}),
\begin{equation}
\mathrm{CODATA:}\ \  \sigma_{\mathrm{th}} =  5.670400(40) \cdot 10^{-8}\ W m^{-2} K^{-4}. \label{codata}
\end{equation}
It is called the theoretical value and has the uncertainty $7 \cdot 10^{-6}$. The value of $\sigma_{\mathrm{th}}$ has not changed since 1998 \cite{CODATA-RMP2000}. Last time the Stefan-Boltzman constant was determined by direct measurements for the 1986 edition of CODATA \cite{CODATA-RMP1987}. 

However, additional  contributions to thermal radiation exist because the Stefan-Boltzmann law is derived  under the assumption   that radiation wavelengths are infinitesimal compared with the geometrical dimensions of a cavity or a body \cite{Sommerfeld-bookV5,Landau-V5} (textbooks let the volume go to infinity, but this limit is not mathematically and physically defined). For arbitrary wavelengths and temperatures, the radiation laws are sensitive to the size and the shape of an emitter. In other words, the Stefan-Boltzman law is only approximate. Under typical experimental conditions in metrology, it can be considered exact only for large enough bodies (cavities), this condition we specify below. 

Different physical quantities can only compared when they are expressed in the same dimensionality, i.e., in the same physical units. As the geometrical parameter of a physical system (body or cavity), one can take its {\em effective size} $r$, which is defined as the ratio of its volume to its boundary's area,
\begin{equation}
 r \equiv \frac{{V}}{{S}},  \label{effsize}
\end{equation}
e.g., the effective size of a sphere of radius $b$ is $r =b/3$. The radiation's properties are characterized by the thermodynamic (absolute) temperature $T$ (Kelvin), which can be expressed  as the scaled inverse temperature (meter), with help of the fundamental constants \cite{Planck-book1914}
\begin{equation}
\beta = \frac{\hbar c}{n T k_{\mathsf{B}} }, \label{beta}
\end{equation}
where $k_{\mathsf{B}}$ is the Boltzmann's constant, $\hbar$ is the Planck's constant, and $c$ is the speed of light, and $n$ is the calibrating number. This Planck's inverse temperature $\beta$, known in the condensed matter theory as the thermal wavelength, should be used in the physics of finite size thermal emitters instead of the absolute temperature $T$. Then, the 'high temperature' limit should be understood as the asymptotic,
\begin{equation}
\frac{\beta}{r} \ll 1. \label{highT}
\end{equation}
Incidentally, the opposite ('low temperature') limit $\beta \gg r$ limit can be reached for any physical system by sufficiently lowering temperature, while reaching the limit (\ref{highT}) may be restricted by thermal destruction.

The integral law of thermal radiation for finite size emitters is usually sought as a series for the radiant exitance (or the free energy) in the small dimensionless parameter (\ref{highT}),
\begin{equation}
j= \sigma_{\mathrm{th}} T^4 \Big(1 + \kappa (rT)^{-1} + \epsilon (rT)^{-2} + \mathrm{O}[{\mathrm{geometry}}, T]\Big). \label{highTj}
\end{equation}
The error term was assumed to be the higher orders of $1/(rT)$, but it really depends on the emitter's geometrical characteristics other than volume and boundary's area.  In fact, the expression above cannot be even considered an expansion in a small  parameter, because it is not derived, but rather assumed. Indeed, to know the convergence of the series (\ref{highTj}) and the behavior of the discarded remainder in (\ref{highTj}) one should know the generating functional, i.e., the radiant exitance $j$, as a functional of arbitrary temperature and arbitrary geometry of an emitter (not only its volume and boundary's area, but also its shape).  Thus, the last term of (\ref{highTj})  can be put in the displayed form only for a sphere. Let us suggest to take the integral law of thermal radiation for finite size emitters as a short expression,
\begin{equation}
j= \sigma_{\mathrm{th}} T^4 \Big(1 + \frac{\kappa}{rT}\Big ), \label{mycorr}
\end{equation}
where two terms in the brackets can have any ratio, while an omitted remainder, for some physical systems and conditions, may be greater than both terms above. The second term in (\ref{mycorr}) comes from the free energy contribution, which is proportional to the boundary area. Its existence, known as the Weyl's conjecture, has been debated for a long time.  

We propose that the parameter $\kappa$ is given by the universal expression,
\begin{equation}
\kappa = \frac{15 \zeta(3)}{4 \pi^3 }\, \frac{\hbar c}{ k_{\mathsf{B}} } 
\approx 3.33 \cdot 10^{-4}\, K m. \label{kappa}
\end{equation}
The functional form of $\kappa$ and, correspondingly, its order of magnitude is obvious from the expression (\ref{beta}) or can be deduced by the dimensional analysis, therefore, the question is only its numerical prefactor. The zeta function value above is $\zeta(3) \approx 1.202$, so the coefficient at $(\hbar c)/k_{\mathsf{B}}$ in (\ref{kappa}) is 0.145 in comparison to 0.5 in the usual choice for the definition (\ref{beta}). Similar forms for the kappa parameter, also based on $\zeta(3)$, were suggested in literature \cite{Maslov-TMF2008,Maslov-MN2008,Baltes-PLA1969,Baltes-OC1970}, by using the volume based length $V^{1/3}$ instead of (\ref{effsize}), but commonly this parameter is taken zero. The expression (\ref{kappa}) is based on the incomplete derivations in the field theory, thus, so far it should be considered as an empirical fit.

The boundary's area contribution to thermal emission (\ref{mycorr}) has not been studied in  theoretical and experimental physics of thermal radiation, because this additional term had been assumed absent. Let us review some preceding theoretical works that assert it is absent. H.P. Baltes and F.K. Kneub\"uhl derived the $T^3$ contributions to the radiation energy \cite{Baltes-PLA1969,Baltes-OC1970}. The correction given in \cite{Baltes-PLA1969} has a negative sign and by an order of magnitude larger than (\ref{kappa}) and than available experimental data. Ref.~\cite{Baltes-OC1970} is based on numerical simulations, which report that the $T^3$ term vanishes for some shapes of the cavity, but not for others. In the following papers \cite{Baltes-OC1971,Baltes-PRA1972,Baltes-AP1973} this term was discarded apparently to comply with  Refs. \cite{Case-PRA1970} and \cite{Balian-AP1971}. In later publications, the absence of the boundary's area term became a commonly accepted fact, and it was no longer questioned or verified, e.g., \cite{Garcia-PRA2008}. However, the main argument of Ref.~\cite{Case-PRA1970} is only a standard textbook's statement that, in the limit of an emitter's infinite size, the boundary's area contribution becomes infinitesimal compared to the volume one. But the size characteristic (\ref{effsize}) never vanishes and it cannot be taken zero in theoretical derivations that investigate the finite size effects of real physical systems. The same paper stated that this absence can be proved by methods of Ref.~\cite{Kac-AMM1966}, however,  \cite{Kac-AMM1966} really tried to show the existence of the boundary area term (although in a mathematically different problem). The paper \cite{Balian-AP1971} stated that the boundary area term disappears due to cancellation of the one-reflection Green's functions contributions. Unfortunately, this method is unacceptable for the Laplacian problems on compact manifolds, where one should seek the global late-time solutions.  Besides, Ref.~\cite{Balian-AP1971} solved the radiation problem in a cavity with perfectly conducting boundaries, while the experiment's blackbody surface is dielectric \cite{Quinn-PTRSLA1985}, p. 98. We aim to describe the universal contributions of the integral law of thermal radiation, while its dependence on the emitter material appear in other contributions and under other physical conditions. Summarizing, the $T^3$ contribution (the boundary area term) in the integral law of thermal radiation was deemed nil, but no proofs were given. The field theory of thermal radiation is yet to be developed.

Let us proceed now to review some experiments and observations in the field of thermal radiation in order to reaffirm the integral law of thermal radiation given by the equation (\ref{mycorr}). This phenomenological analysis is used to address the following questions: 1) existence of the boundary area contribution in the total radiation flux, 2) the order of magnitude of this contribution as defined by $\kappa$, 3) the relative ratio of two contributions, due to volume and to boundary's area, in different $r$ vs. $T$ regimes.

\section{Experiments} \label{experiments}

\subsection{The radiometric determination of 
the Stefan-Boltzmann constant at National Physical Laboratory} \label{SBconstant}

An overview of the early history of measurements of the Stefan-Boltzmann constant is given in \cite{Quinn-PTRSLA1985}). The first reported deviation from the Stefan-Boltzmann law under then standard radiometric conditions was about $1.8\%$ \cite{Gillham-PRSA1962}. It persisted for a few decades, and it is much bigger than the deviation (\ref{delta}) or \cite{Maslov-TMF2008,Maslov-MN2008}.  Apparently, that deviation was some experimental artifact due to the older method of J. Guild \cite{Guild-PRSA1937}. When the new kind of experiment was performed decades later \cite{Blevin-Metro1971} it gave the first modern value of $\sigma$. 

The most precise, direct determinations of the Stefan-Boltzman constant by radiometric observations were done in 1980's at National Physical Laboratory, U.K. by T.J. Quinn and J. Martin. They have designed and built an experimental device that detected infrared radiation from a blackbody cavity by a cryogenic radiometer \cite{Quinn-PTRSLA1985}.  The  NPL team posed a question \cite{Quinn-PTRSLA1985}: {\em "Haw large must a cavity at 233 K be for the Stefan-Boltzmann formula to represent correctly the total radiant energy density to better than one part in $10^5$?"} The geometrical sizes taken from the figure (since no numbers are given) in \cite{Quinn-PTRSLA1985,Parr-book2005} specify the NPL blackbody radiator as a cylinder with the height $32\ cm$ and the diameter  $11\ cm$. At the bottom, it has a cone of the height $21\ cm$ that increases the total surface and decreases the volume of the radiator. A straightforward computation of volumes and areas of the cylinder and the cone gives the radiator's effective size as $r = 1.6\ cm$. There might be some small discrepancy with actual numbers because we used of a ruler and a derived scale, and neglected the radiator's aperture.

To answer the question above the NPL study used theoretical derivations of H.P. Baltes \cite{Baltes-PRA1972,Baltes-AP1973} who directly summed the  spectra of electromagnetic radiation inside a cylinder as an approximation to the NPL radiator. The energy formula in \cite{Quinn-PTRSLA1985} obtained with help of Baltes' results assumes a series in the powers of $1/(r T)$. It contains corrections to the Stefan-Boltzmann law starting from the second order. These and higher order terms are defined by the shape and edges of the radiator. Similar corrections were derived in \cite{Pathria-PRA1973} by the methods of spectral theory. The introduced uncertainty was estimated to be of order $10^{-7}$ (or $0.01\ mK$) \cite{Quinn-PTRSLA1985}. We suppose that the missing linear term, $\kappa/(rT)$, accounts for much of the experiment's uncertainty.   

The mean value of the Stefan-Boltzmann constant obtained by the NPL experiment is \cite{Quinn-PTRSLA1985} (scaled here as in \cite{CODATA-RMP1987}),
\begin{equation}
\mathrm{NPL:}\  \sigma_{\mathrm{exp}} =  5.66967(76) \cdot 10^{-8} W m^{-2} K^{-4},
\end{equation}
with the uncertainty, 
\begin{equation}
\delta_{\mathrm{exp}} = 1.4 \cdot 10^{-4}. \label{NPLerror}
\end{equation}
Theoretically, the intrinsic uncertainty of the radiating power from a thermal source introduced by the $T^3$-contribution in (\ref{mycorr}) is,
\begin{equation}
\delta = \frac{\kappa}{r T},  \label{delta}
\end{equation}
or it is $(\delta \cdot 100)\%$. For the NPL determination temperature, $T=273\ K$ this formula gives $\delta \approx 0.76 \cdot 10^{-4}$. Accepting that only half the value (\ref{NPLerror}) comes from the blackbody radiator \cite{Quinn-PTRSLA1985}, this is a fairly good agreement with (\ref{delta}). In principle, Eq.~(\ref{delta}) could be used to match the observable $\delta_{\mathrm{exp}}$ in order to deduce the $\kappa$-factor (\ref{kappa}).  

The total radiation thermometer and the CODATA value of the Stefan-Boltzmann constant (\ref{codata}) could be used to measure the radiant exitance as a function of temperature (\ref{mycorr}). However, it is impossible to do with the data of \cite{Quinn-PTRSLA1985}, because all measurements of $\sigma$ were done in the close proximity of a single temperature value, triple point of water. NPL has planned to achieve the measurement of $\sigma$ with uncertainty of $4 \cdot 10^{-5}$, but since 1986 the determination of the Stefan-Boltzman constant is no longer done by the radiometer technique \cite{CODATA-RMP2012}. With the given design and the temperature range, the blackbody's intrinsic uncertainty due to the cubic term in the integral law of thermal radiation would have limited the NPL experiment's precision by the uncertainty $10^{-4}$, thereby, preventing further progress. 

In another work \cite{Martin-Metro1988}, with the same experimental setup, the cryogenic radiometer was used to measure the blackbody cavity's temperature, which, together with \cite{Quinn-PTRSLA1985}, covers the range from $-130$ to $+102\,^{\circ}$C. These data for the radiant temperature (\ref{Tsb}) can be compared with the thermodynamic (contact) temperature as measured by the platinum resistance thermometers installed in the blackbody cavity. The radiant exitance, measured by the radiometer in work \cite{Martin-Metro1988}, was assumed to obey the Stefan-Boltzmann law (\ref{Tsb}) with a corresponding radiant temperature $T_{\mathsf{SB}}$. This power should be equal to the one derived the extended formula (\ref{mycorr}). Solving this equation gives the corrected  radiant temperature, 
\begin{equation}
T_{\mathsf{SB}} \approx T \Big( 1 + \frac{\kappa}{4 r T} \Big), \label{Tcorr1}
\end{equation}
which is, therefore, always higher than the true temperature $T$, as measured by the resistance thermometers. As follows from (\ref{Tcorr1}), this difference should be a constant defined by the effective size, 
\begin{equation}
	\Delta T = (T_{\mathrm{SB}} -T) \approx \frac{\kappa}{4 r}, \label{deltaT}
\end{equation}
and  for the NPL experimental device it is about $5\ mK$.  

From the data reproduced in the table of \cite{Martin-Metro1988}, we extracted the  difference between the thermodynamic and radiometric temperatures.  This difference does not appear to obey Eq.~(\ref{deltaT}). To clarify the problem, let us first deal with the thermometer's calibration.  The thermometers were calibrated with the International Practical Temperature Scale of 1968 (IPTS-68) \cite{Ward-Metro1979}.  The new standard, International Temperature Scale of 1990 (ITS-90) \cite{Preston-Metro1990}, indicates significant differences from the previous standard's scale. As can be seen from a simple comparison with the calibrating figures in \cite{Martin-Metro1988} and \cite{Preston-Metro1990}, these differences can approximately account, in the signs and in the values, for the observed divergence from the law (\ref{deltaT}). Quantitatively, we used the polynomial representation of the difference between ITS-90 and IPTS-68 from $83.8\ K$ to $903.8\ K$ ($630.6\,^{\circ}$C) derived by R.L. Rusby and given in the BIPM report \cite{Fellmuth-BIPM2012} to obtain  $(T-T_{68})-(T_{90}-T_{68})=T-T_{90}=\Delta T$. The mean of the computed $\Delta T$ is statistically significant zero. This is not surprising, because this treatment did not remove all artifacts of the IPTS-68 calibration: the corrected $\Delta T$'s are still positive below $0\,^{\circ} C$ and negative above. Besides, the experiment's uncertainties go up to $4.3\ mK$  \cite{Martin-Metro1988}, comparable to the theoretical value of $\Delta T$, which makes it difficult or impossible to confirm the formula (\ref{deltaT}). A different experimental setup discussed in the next section is capable of detecting (\ref{deltaT}), so it could produce the kappa factor (\ref{kappa}), which is no less important than the Stefan-Boltzmann constant. 

\subsection{The calibration of cryogenic blackbodies at  
National Institute of Standards and Technology} \label{NIST}

Let us address the problem with unexplained uncertainties  in the power (temperature) determination during the calibration of cryogenic blackbodies performed in the Low Background Infrared Calibration Facility (LBIR) at the National Institute of Standards and Technology (NIST), USA. This facility was built to develop sensitive infrared detectors as a part of the 'star wars' program \cite{Mekhontsev-book2010}. For this purpose, blackbodies (cavities) with very low radiance were produced. Geometrical characteristics of these blackbodies are not disclosed due to the confidential nature of these blackbody calibrations \cite{Carter-Metro2006}. One can still use some experimental data from the radiation measurements that were published in \cite{Datla-NIST1994} and essentially reprinted in \cite{Parr-book2005}. This is it, we do not need to know an effective size of the tested blackbody in order to derive functions of temperature (\ref{mycorr}) and (\ref{deltaT}). It is sufficient to know that $r$ is constant for a given tested blackbody and treat it as a sought unknown parameter.

The contact temperature from one resistance thermometer, $T_1$ is in the first column of Table \ref{NISTdata}. It gives the radiant power, $j_{\mathsf{SB}}$ calculated according to the Stefan-Boltzmann law, in the third column. The measured average power, $\Phi$ was given for the range of blackbody's temperatures from 200 to 400 K in Table 4 of \cite{Datla-NIST1994}. Geometrical characteristics of the experimental setup  (radii of both apertures, distance) from that paper boil down to $A= 3.31 \cdot 10^{-7}\ m^2$ and $B= 2.36 \cdot 10^{-3}$ that enter the formula (\ref{jPhi}). This allows us to derive from $\Phi$ the radiant exitance, $j$, which is in the second column of Table \ref{NISTdata}.  The difference between $j_{\mathsf{SB}}$ and $j$ is in the fourth column. The radiant temperature $T_{\mathsf{SB}}$ by Eq.~(\ref{Tsb}) is in the fifth column. Finally, the difference between these temperatures $\Delta T$ is the last column.
\begin{table*}[!h]
\caption{The NIST calibration of cryogenic blackbodies (based on \cite{Datla-NIST1994})}
\label{NISTdata}
\begin{tabular}{llllll}
\hline
$T_1 (K)$ & $j (W/m^2)$  & 
$j_{\mathsf{SB}} (W/m^2)$&  $\Delta j (W/m^2)$  & $T_{\mathsf{SB}} (K)$ & $\Delta T (K)$ \\
\hline
199.90 &  92.73&  90.56 & 2.17 & 201.09 &  1.09 \\
224.78 & 146.78 & 144.75 & 2.02 & 225.56 & 0.56 \\
249.68 & 222.72 & 220.36 & 2.36 & 250.34 & 0.34 \\
274.68 & 326.59 & 322.81 & 3.79 & 275.49 & 0.49 \\
299.55 & 462.61 & 456.53 & 6.08 & 300.54 & 0.54 \\
324.43 & 638.59 & 628.19 & 10.40 & 325.76 & 0.76 \\
349.36 & 857.86 & 844.73 & 13.13 & 350.71 & 0.71 \\
374.12 & 1132.84 & 1110.86 & 21.98 & 375.96 & 0.96 \\
399.07 & 1465.96 & 1438.18 & 27.78 & 400.98 & 0.98
\\\hline
\end{tabular}
\end{table*}

By analyzing these data, we test a hypothesis that the radiant exitance is a general function,
\begin{equation}
j = \sigma_{\mathrm{th}} T^4 + \lambda T^3, \label{nist}
\end{equation}
with the given $\sigma_{\mathrm{th}}$ and the unknown constant $\lambda$. It is obvious from the fourth column, that a  series for the power excess, $\Delta j$, accepts a fit by the cubic function of temperature,
\begin{equation}
\Delta j = \lambda T^3. \label{cubic}
\end{equation}
The chi-square test of statistical significance of the fitting function (\ref{cubic}) with the obtained parameter,
\begin{equation}
 \lambda = 3.63(30) \cdot 10^{-7}\ W/(m^2 K^3), \label{lambda}
\end{equation}
cannot reject the null hypothesis of this cubic law. Assuming the first two numbers are an experimental artifact and discarding them does not change much the result. Here we first derived $\Delta j$ and then fit it, but similarly the whole function (\ref{nist}) can be successfully fit to the data from the second column, in which case $\sigma_{\mathrm{exp}}$ can be also obtained. The difference between the thermodynamic and radiant temperatures, $\Delta T$ in the last column, accepts a statistically significant fit by the constant $\Delta T = 0.715(85)\ K$, thereby confirming (\ref{deltaT}). Using the numerical values (\ref{codata}) and (\ref{kappa}), together with (\ref{lambda}) for $\lambda= {\sigma_{\mathrm{th}}\kappa}/r$, we obtain the effective size $r=5.1 \cdot 10^{-5}\ m$ for a tested blackbody. This number can only be verified by the NIST authors, but it is consistent with the given size of a blackbody's aperture.

However, this is not a full answer yet. Ref.~\cite{Datla-NIST1994} shows that $\Delta T$, computed with the average of two thermometers temperatures, is linearly proportional in temperature. This fact is further confirmed (without displaying any numerical data though) in the figures and the text of \cite{Carter-Metro2006}. If this feature is not an experimental or calibration artifact, then it could only be an effect of higher order terms in the power. Since they were discarded as negligible in \cite{Quinn-PTRSLA1985}, quite likely they were neglected in \cite{Datla-NIST1994} as well. This guess is supported by the fact that the effective size $r$ computed from the temperature difference (\ref{deltaT}) is twice as big as found above. The next order in the expansion by (\ref{beta}) for the given values of $r, \kappa, T$ is small, but not negligible (while it was in \cite{Quinn-PTRSLA1985}). Then, the additional $T^2$ contribution to the radiant power would result in a temperature dependence of (\ref{deltaT}). This is the effect of smallness of a cavity: when the Planck's inverse temperature (\ref{beta}) and the effective size (\ref{effsize}) become comparable,  the condition (\ref{highT}) weakens.

The phenomenon of an excess of the radiation power was found twenty years ago. It is well measured but its nature remains undetermined \cite{Mekhontsev-book2010}. {\em "The various sources of uncertainty were covered in detail to show that experimental error is not capable of explaining the temperature errors observed"}, \cite{Carter-Metro2006}. We believe that the experimental observations in Refs.~\cite{Datla-NIST1994,Parr-book2005} confirm, in the first approximation, the cubic addition to the Stefan-Boltzmann law (\ref{mycorr}) and the constant law for the difference between the radiant and thermodynamic temperatures (\ref{deltaT}). 

\subsection{The size of source effect and blackbodies in metrology} \label{SSE}

The surface term in the thermal radiation flux makes it dependent on the source's geometrical properties. In metrology the uncertainty (error) due to the source's geometrical characteristics is well known, it is called the size-of-source effect (SSE). It is assumed that the radiation detector picks up the energy flux from outside the area it is aimed at: {\em ``Due to a non null sensitivity of radiance measurement device outside the target field we must subtract from the signal the amount of flux coming from the surrounding part of the source''} \cite{BIPM-2008}. We conjecture that a significant part of the SSE comes from the excess of radiance over the Stefan-Boltzmann formula, and it cannot be and should not be removed.  The analysis of the size-of-source effect in the monograph  \cite{Quinn-book1990} seems to confirm  this hypothesis. First, theoretical derivations of the diffraction's contribution give 2-4 times less power than the measured radiant flux of SSE. When the diffraction effects are properly accounted for, what other physical processes could transfer extra energy? Second, this effect is observed only when comparing tungsten lamps with blackbodies in furnace. These calibrations are analyzed below. 

Let us focus on the study done in 2008 by Bureau International des Poids et Mesures (BIPM). BIPM undertook the cross-comparison (calibration) of  standard light sources from the world's main metrology  laboratories (NIST, NRC, PTB, VNIIOFI, BNM-INM) \cite{BIPM-2008}. This study measured and compared the spectral radiance of tungsten lamps.  The above formulae give no information about spectra but the boundary area contribution varies with the temperature and source's size (\ref{effsize}), so the deviation from the Stefan-Boltzmann law (\ref{delta}) varies accordingly. The lamps (lightbulbs) in the BIPM study have tungsten ribbon filaments as a source of radiation. The spectrum was split into two spectral zones, and the range of wavelengths from $300\ nm$ to $1050\ nm$ was calibrated with the pyrometric standard lamp (Lpyr) used as a blackbody. For the wavelengths from $950\ nm$ to $2500\ nm$, a copper fixed point blackbody (CuBB) was used.  So, in the first spectral range lamps are compared with lamps, while in the second one, lamps compared with CuBB. The hypothesis is that this is the origin of a larger uncertainty in the latter case. 

In the first spectral zone, the study assumed (Table 3.6 of \cite{BIPM-2008}) no uncertainty due the size of source effect. The geometrical sizes of the lamps' filaments, probably including Lpyr, are similar and, therefore, their effective sizes are similar as well. Thus, comparing radiant powers (radiant temperatures) of the lamps should show no boundary energy contributions (no size-of-source effect) if a thermometer were also calibrated with a lamp blackbody. This means that an intrinsic extra power (\ref{delta}) or temperature (\ref{deltaT}) would be assigned as an uncertainty to other origin, in this case, to the CuBB blackbodies. 

In the second wavelength's range, CuBBs were used. A typical copper fixed point blackbody presents a radiating cavity in the copper cylinder (ingot) at the copper freezing temperature $T_{CuBB}=1357\ K$ \cite{NIST-CoBB}. From the description of the device, an estimate (assuming these are typical sizes used in this study) for the CuBB's effective size is $2.6\ mm$. Formula (\ref{deltaT}) gives for a CuBB with these  characteristics  the uncertainty in temperature $2.3 \cdot 10^{-5}$ or $0.002\%$. This contrasts with value $0.08\%$  in Table 3.6 of \cite{BIPM-2008} , but agrees by the order of magnitude with $0.0087\%$ in  Table 2.1. Both tables are calibrations with CuBBs, so more technical information is needed to clarify how an uncertainty gets propagated in the calibrations. However, its is clear that a CuBB blackbody can be accepted as an ideal, Stefan-Boltzmann, blackbody (\ref{Tsb}). 

Reported geometrical characteristics of the tested tungsten lamps included only the length and the width of their filament ribbons, but not their thickness. The only lamp with the declared thickness was used at  Bureau National de M\'etrologie (France) \cite{BIPM-2008}: 'Polaron 24/G/UV' product has its filament thickness $l=0.07\ mm$ \cite{Polaron}. The effective size of the filament made as a thin ribbon can be approximated as $r \approx l/2$, i.e.,  the filament's effective size is completely defined by its thickness.  For a lamp at temperature $1910\ K$, the temperature uncertainty caused by the boundary energy is $1.2\cdot 10^{-3}$ or $0.12\%$. The accepted in this BIPM study uncertainty was $0.30\%$ (Table 3.7, \cite{BIPM-2008}). This discrepancy might be explained by the fact that the tungsten filament's surface, heated by DC currents, develops grooves and facets \cite{Quinn-book1990}. This phenomenon can increase the  area, while keeping the volume  approximately constant, which would result in a decreased effective size and a higher power and higher uncertainty. Recent experiments on thermal emission controlled by the lab produced gratings on tungsten surfaces \cite{Han-OC2013} may be a way to validate this idea.

The validity of the Stefan-Boltzmann law is limited to the asymptotic (\ref{highT}). This approximation (\ref{Tsb}) used to be sufficient for measurements, but in the last two decades technology demanded a better precision. In response, empirical corrections were suggested \cite{Bloembergen-Metro2009}, but those could not be universal, 
{\em ``When applying corrections, it is necessary, in principle, to recognize which SSE quantity has been measured and under which conditions, as the correction formulae for each quantity differ and may also be functions of the measurement parameters''}, \cite{Sounders-IJT2011}. The general formulae (\ref{delta}) and (\ref{deltaT}) are proposed for testing. It is also easy to derive a correction formula for the physical (thermodynamic) temperature $T$ of a radiating source from the observed (radiant) temperature $T_{\mathsf{SB}}$, as measured by a radiometric device that operates on the assumption of  validity of the Stefan-Boltzmann law (\ref{Tsb}). From knowing the emitter's effective size, when the deviation is small enough (the condition normally valid in radiometry), it is,
\begin{equation}
T \approx T_{\mathsf{SB}} \Big( 1 - \frac{\kappa}{4 r T_{\mathsf{SB}}} \Big). \label{Tcorr}
\end{equation}

This conjecture changes the interpretation of the (part of) size of source effect, so that it can only only observed when comparing light sources of different kinds (effective sizes), or when measuring temperature of sources of one kind by the thermometers calibrated with blackbodies of another kind. Since an uncertainty does not have a sign, the traditional interpretation of SSE cannot indicate whether an extra power comes from CuBBs or lamps. It is suggested here that  a radiometer detects the higher radiance from a tungsten lamp than from a furnace blackbody, at the same thermodynamic temperature, but this effect is interpreted as a (part of) SSE of a furnace blackbody due to the thermometer's calibration.

As an experimental supporting evidence, let us consider a recent work on the enhanced thermal emission from antenna-like nanoheaters \cite{Ingvarsson-OC2007}, whose physical parameters are reasonably close to the incandescent lamps discussed above. Indeed, the platinum wires were made on a silicon substrate, with the length $6\, \mu m$, the thickness  $40\, nm$ and the width ranging from $4$ to $0.125\ \mu m$. The polarized thermal emission (polarization of thermal emission by thin wires is studied experimentally in more detail in \cite{Bimonte-NJP2009}) was measured at temperatures $180 \ldots 650\,^{\circ}$C, created by the applied voltage.  The total radiation efficiency was obtained by a normalization procedure, which amounts to discarding the overall factor in (\ref{mycorr}). The results were given in Plot 3 \cite{Ingvarsson-OC2007} in arbitrary units for the radiation efficiency. Displaying the plot flat at zero efficiency amounts to the subtraction of 1 in the brackets of (\ref{mycorr}), so, this plot can be viewed merely as a function of $\delta$ (\ref{delta}). The sharp increase in this plot is observed with decreasing the heater's width from $0.5$ to $0.125\, \mu m$. It means the radiance is several times greater than the Stefan-Boltzmann law. The functional behavior $r^{-n}$, with $n \ge 1$ ($n=1$ in the present work) can be approximately seen from the displayed data points. Further analysis requires the experiment's data, but for correct comparison the nanoheater should be suspended, not on a substrate. An analytical method, based on the theory of Sergei Rytov \cite{Rytov-book1978}, for thermal radiation spectra of long cylinders (wires) is being developed \cite{Kardar-PRE2012}. Integrating its spectra should allow for comparison with the nano-heaters experiment and the  equation proposed in our paper.  

Considering the growing number of experimental optics works performed in the infrared range of frequencies \cite{Schuller-NPhot2009,Han-OC2013,Ingvarsson-OC2007}, these estimates would hopefully motivate detailed analyses.  This proposal agrees with currently accepted uncertainties (around $1\%$) in calibrations of the NIST optical standards \cite{Parr-NIST2007}. The direct way to test it is to compare blackbodies or standard light sources of the same type but of different effective sizes. Therefore, the following experimental verification can be suggested.  The metrology labs could produce the copper fixed point blackbodies of different shapes in order to modify the effective size ${V}/{S}$. The larger difference in effective sizes, the higher uncertainty (\ref{deltaT}) in the calibration of CuBBs.

\section{Discussion} \label{discussion}

Above we analyzed experimental works on thermal radiation to investigate questions of the existence and magnitude of a contribution to the radiance due the boundary of a physical system. Two metrology experiments were chosen because they represent a state of the art in precision experiments. The analysis of the NPL data showed that it is not possible to confirm the law (\ref{mycorr}), but theoretical and experimental uncertainties agree. It is possible to get the confirmation from the calibration of cryogenic blackbodies done at NIST. Using more recent data and blackbodies characteristics could give a better fit, but the lack of published data requires verification by the NIST team. The $\kappa$-factor (\ref{kappa}) could be measured in current or future radiometric experiments and the conjecture about its universal nature studied.

From a theorist's point of view the NPL experiment paper \cite{Quinn-PTRSLA1985} and the NIST report \cite{Datla-NIST1994} are excellent examples of the way experimental studies and their data should be presented, with all the details (except no sizes of blackbodies were given). In the cases studied in Sects. \ref{SBconstant} and \ref{NIST}, observables were the power measured by radiometers and resistance thermometers.  These were relatively small tables, but observables and raw data from larger studies should be deposited online. The data for observables are crucially important for the progress of physics, but unfortunately they are often missing from published experimental studies or substituted by predetermination theoretical analyses.

Even though Eq.~(\ref{mycorr}) can show how much an observed radiant exitance deviates from the Stefan-Boltzmann law, this formula is valid for arbitrary effective size $r$ and temperature $T$. However, it ceases to be valid in the asymptotic $r T \ll \kappa$,  because the omitted remainder, which is not universal, can become overwhelming. Therefore, the integral law of thermal radiation still has to be completed to cover these physical conditions. It is clear that when deriving the total radiance of finite size emitters from spectral distributions theoretically obtained in the studies of thermal radiation as an electromagnetic phenomenon, one should neither expect arriving at the Stefan-Boltzmann law, nor normalize by it. The result of such calculations will be new physics beyond Eq.~(\ref{mycorr}). 

Victor Maslov derived \cite{Maslov-TMF2008,Maslov-MN2008} the free energy expression containing a similar contribution, by different methods. The boundary area coefficient in $j$ derived from this free energy  differs in form from (\ref{kappa}) because his theory is based on the characteristic area parameter $V^{2/3}$ vs. $S$. Apparently due to accumulated numerical coefficients, the size of that surface correction is about an order of magnitude larger than the conjectured (\ref{mycorr}) and the one seen in experiments. Nevertheless, it is quite important that the free energy with the boundary area term can be derived in statistical physics, since it is based on  probability theory \cite{Maslov-RJMP2007} and combinatorics \cite{Erdosh-BAMS1946}, while this paper's approach is based on the heat kernel method \cite{CPT2}. The physical meaning of such free energy is, of course, almost universal in physics. It should also be mentioned that early derivations of the boundary effects in statistical physics were done by H.K. Pathria \cite{Pathria-PRA1973,Pathria-book1996}.

It is well known that the origin of the fourth power $T^4$ in the Stefan-Boltzmann law is the dimension four of our spacetime, e.g., \cite{Landsberg-JPA1989}. In fact, the main physical information about the Stefan-Boltzmann law, its fourth power in temperature, was derived by Jo\v{z}ef Stefan \cite{Stefan-1879} from the experiments of P.L. Dulong and A.T. Petit empirically. He was able to do this by introducing the absolute temperature scale \cite{Crepeau-ETFS2007,Dougal-PhysEd1979}. From knowing the fourth power, a combination of physical constants, which comes a factor, is fixed by the dimensionality (\ref{codata}). This is a well known method of the dimensional analysis, e.g. \cite{Migdal-book1977}. The remaining numerical coefficient of the order of unity is to be found from experiments.  Although  the expression (\ref{mycorr}) contains two independent constants for two equally important contributions, similar qualitative reasoning can still be applied. Following this approach, the expression (\ref{mycorr}) can be derived empirically, without referring to any theory. It can simply be used as an {\em ansatz} for the calibration by the radiometry measurements, e.g., Sect. \ref{NIST}, if geometrical characteristics of blackbodies were known. 

An ideal cavity for the radiometer is a sphere, because any other shape would result in a larger area of the cavity's surface with the same volume and, correspondingly, in a higher intrinsic uncertainty (\ref{delta}).  The expression (\ref{mycorr}) becomes the Stefan-Boltzmann law only in the 'large' sphere limit $r \gg \beta$. At temperatures of the order of $10^{-3} K$,  the surface 'corection' can become larger than the leading Stefan-Boltzmann term even for the NPL radiator. These physical conditions are not realized there, but certainly they may occur. The ideal, Stefan-Boltzmann, radiating bodies or cavities do not exist in Nature, because the boundary's area energy remains for any real physical system due its finite size.

The surface curvatures  and edges terms add to the total radiance, but they are infinitesimal in the considered radiometric measurements. In general, all finite size effects were previously considered uncertainties and eliminated through theoretical analysis and experimental improvements. However, removing any of these contributions from the free energy, and correspondingly from the radiance, is not justified from the physical point of view. They are simply physical manifestations of unaccounted terms in the total free energy, and they are not necessarily small. If the Stefan-Boltzmann law is only one term in a general functional of temperature that depends on the matter and geometry of an emitter, then the Stefan-Boltzman constant is not fundamental. Indeed, BIPM already declared it \cite{CODATA-RMP2012} as {\em defined} through the fundamental constants (\ref{sigma}). Therefore, the Stefan-Boltzman constant's use is limited to some practical tasks in the limit (\ref{highT}). Regarding the fundamental constants of physics, they will soon be assigned exact numerical values, since BIPM is about to introduce the {\em new SI} \cite{Becker-Metro2007,BIPM-newSI2011}. This change would realize the dream of theoretical physics, the system of natural units, proposed  by Max Planck in his 1897 lectures \cite{Planck-AdP1900,Planck-book1914}.

The phenomenon discussed in this paper suggests several interesting research directions, because the Stefan-Boltzmann law is assumed to be exact in various areas of physics. One of important applications is the analysis of dusts in fusion plasmas. The dust, i.e. matter particles of nano-micrometer sizes, appears in plasma typically due to interaction of high temperature plasma with the tokomak's walls. The physics of dusty plasma inevitably involves energy transfer due to thermal radiation, which is assumed to obey the Stefan-Boltzmannn law \cite{Shukla-book2002}. However, when the size of particles is comparable with the Planck's inverse temperature (\ref{beta}), the equation (\ref{mycorr}) should be used. For the dust particles in plasma, the average sizes of particles range from 0.1 to 10 $\mu m$, while the particles temperature grows from $10^2$ to $10^4 K$ \cite{Bacharis-PPlasmas2010}. A quick estimate by  (\ref{delta}) shows that the surface power component of thermal radiation for some combinations of these parameters may be as big as and bigger than the Stefan-Boltzmann component. At a later stage of heating, ablation of the dust particles may bring the particle's size down even further until it presumably vanishes. In this regime, dust particles emit as the $T^3$ radiance. This physics should be similar in some aspects to physics of ordinary flame. Perhaps, this unaccounted component of the equations of state, which are built into the computer codes \cite{Bacharis-PPlasmas2010}, may help solving some problems in physics of fusion plasma.


\section*{Acknowledgment}
The author would like to thank CERN Theory Division, Perimeter Institute for Theoretical Physics and Yukawa Institute for Theoretical Physics for hospitality and support during the corresponding visits. Special thanks are to Veselin Jungic at Simon Fraser University's IRMACS Centre for the electronic information support. 


\section*{Postprint note}

The first version of the present paper was submitted to a journal on Oct. 28, 2013. Therefore, this paper cites only the works that were published before that date.


\begin{thebibliography}{99}
\bibitem{Siegel-book1992}
	R. Siegel and J.R. Howell,
	{\em Thermal radiation heat transfer}, third ed.
	(Hemisphere Publishing Co, Washington DC, 1992).
\bibitem{Sommerfeld-bookV5}
	A. Sommerfeld, 
	{\em Lectures on theoretical physics, Volume V. 
	Thermodynamics and statistical mechanics}
	(Academic Press, New York, 1956), sect. 20.
\bibitem{Landau-V5}
	L.D. Landau and E.M. Lifshitz,
	{\em Course of theoretical physics. Volume 5. Statistical physics. Part 1.} 
	(Pergamon, Oxford, 1980), sect. 63.
\bibitem{BaltesHilf-book1976}
	H.P. Baltes and E.R. Hilf,
	{\em Spectra of finite systems. A review of Weyl's problem.}
	(Zurich: Bibliographisches Institut, 1976).
\bibitem{Garcia-PRA2008}
	A.M. Garc\'ia-Garc\'ia,
	Finite-size corrections to the blackbody radiation laws.
	Phys. Rev. A {\bf 78} 023806(5) (2008).
	\href{http://dx.doi.org/1050-2947/2008/78(2)023806(5)}{doi:1050-2947/2008/78(2)023806(5)}
\bibitem{Weyl-JRAM1913}
	H. Weyl,
	\"Uber die Randwertaufgabe der Strahlungstheorie und asymptotische Spektralgeometrie. 
	[About the boundary value problem of the radiation theory 
	and the asymptotic spectral geometry].
	Journal f\"ur die reine und angewandte Mathematik (Crelle's journal),
	{\bf 143}, 177-202 (1913).
	\href{http://dx.doi.org/10.1515/crll.1913.143.177}{doi:10.1515/crll.1913.143.177}
\bibitem{Weyl-MA1912}
	H. Weyl,
	Das asymptotische Verteilungsgesetz der Eigenwerte linearer partieller
	Differentialgleichungen (mit einer Anwendung auf die Theorie der Hohlraumstrahlung).
	[The asymptotic law of distribution of the eigenvalues of linear partial differential equations (with an application to the theory of blackbody radiation)]. 
	Mathematische Annalen {\bf 71}, 441-479 (1912). Available at
	\href{http://www.emis.de/cgi-bin/jfmen/MATH/JFM/quick.html?first=1&maxdocs=20&type=html&an=JFM\%2043.0436.01&format=complete}{www.emis.de}.
\bibitem{Maslov-TMF2008}
	V.P. Maslov,
	Quasithermodynamic correction to the Stefan-Boltzmann law.
	Theor. Mathem. Phys. {\bf 154}, 175-176 (2008).
	\href{http://dx.doi.org/10.1007/s11232-008-0015-x}{doi:10.1007/s11232-008-0015-x}
\bibitem{Maslov-MN2008}
	V.P. Maslov,
	Quasithermodynamics and a correction to the Stefan-Boltzmann law.
	Math. Notes {\bf 83}, 72-79 (2008).
	\href{http://dx.doi.org/10.1134/S0001434608010094}{doi:10.1134/S0001434608010094}
\bibitem{Schuller-NPhot2009}
	J.A. Schuller, T. Taubner, M.L. Brongersma,
	Optical antenna thermal emitters.
	Nature Photonics {\bf 3}, 658-661 (2009).
	\href{http://dx.doi.org/10.1038/nphoton.2009.188}{doi:10.1038/nphoton.2009.188}
\bibitem{Greffet-Nature2002}
	J.-J. Greffet, R. Carminati, K. Joulain, J.-P. Mulet, S. Mainguy, and Y. Chen,
	Coherent emission of light by thermal sources.
	Nature {\bf 416}, 61-64 (2002).
	\href{http://dx.doi.org/10.1038/416061a}{doi:10.1038/416061a}
\bibitem{Fisenko-JPDA1999}
	A.I. Fisenko and S.N. Ivashov,
	Determination of the true temperature of emitted radiation 
	bodies from generalized Wien's displacement law.
	J. Phys. D: Appl. Phys. {\bf 32}, 2882-2885 (1999).
	\href{http://dx.doi.org/0022-3727/32/22/309}{doi:0022-3727/32/22/309}
\bibitem{Fischer-BIPM2008}
	Bureau International des Poids et Mesures (BIPM),
	Uncertainty budgets for calibration of radiation thermometers below the silver point. 
	CCT-WG5 on radiation thermometry (BIPM, Sèvres, France, 2008).
	Available at \href{http://www.bipm.org/wg/CCT/CCT-WG5/Allowed/Miscellaneous/Low_T_Uncertainty_Paper_Version_1.71.pdf}{www.bipm.org/Low T\_{}Uncertainty\_{}Paper\_{}Version\_{}1.71.pdf}
\bibitem{Sounders-Metro2003}
	P. Saunders and D.R. White,
	Physical basis of interpolation equations for radiation thermometry.
	Metrologia {\bf 40}, 195-203 (2003).
	\href{http://dx.doi.org/0026-1394/40/4/309}{doi:0026-1394/40/4/309}
\bibitem{Carter-Metro2006}
	A.C. Carter, R.U. Datla, T.M. Jung, A.W. Smith, J.A. Fedchak,
	Low-background temperature calibration of infrared blackbodies.
	{Metrologia}  {\bf 43}, S46-S50 (2006).
	\href{http://dx.doi.org/10.1088/0026-1394/43/2/S10}{doi:10.1088/0026-1394/43/2/S10}
\bibitem{Ingvarsson-OC2007}
	S. Ingvarsson, L.J. Klein, Y.-Y. Au, J. A. Lacey, and H. F. Hamann,
	Enhanced thermal emission from individual antenna-like nanoheaters.
	Optics Express {\bf 15}, 11249-11254 (2007).
	\href{http://dx.doi.org/10.1364/OE.15.011249}{doi:10.1364/OE.15.011249}
\bibitem{Parr-book2005}
	A.C. Parr and R.U. Datla (eds.),
   {\em Optical radiometry}
	(Elsevier, Amsterdam, 2005).
\bibitem{CODATA-RMP2012}
	CODATA, prepared by P.J. Mohr, B.N. Taylor, D.B. Newell,
	CODATA recommended values of the fundamental physical constants: 2010.
	{Rev. Mod. Phys.} {\bf 84}, 1527-1605 (2012).
	\href{http://dx.doi.org/10.1103/RevModPhys.84.1527}{doi:10.1103/RevModPhys.84.1527},
	\href{http://arxiv.org/abs/1203.5425}{arXiv:1203.5425}	
\bibitem{CODATA-RMP2000}
	CODATA, prepared by P.J. Mohr and  B.N. Taylor,
	CODATA recommended values of the fundamental physical constants: 1998.	
	{Rev. Mod. Phys.} {\bf 72}, 351-495 (2000).
	\href{http://dx.doi.org/10.1103/RevModPhys.72.351}{doi:10.1103/RevModPhys.72.351}
\bibitem{CODATA-RMP1987}
	CODATA, prepared by E.R. Cohen, B.N. Taylor,
	The 1986 adjustment of the fundamental physical constants.
	{Rev. Mod. Phys.} {\bf 59}, 1121-1148 (1987).
	\href{http://dx.doi.org/10.1103/RevModPhys.59.1121}{doi:10.1103/RevModPhys.59.1121}
\bibitem{Planck-book1914}
	M. Planck and M. Masius, 
	{\em The theory of heat radiation}
	(Blakinston's,  Philadelphia, 1914), sect. 164.
\bibitem{CPT2}
	A.O. Barvinsky and G.A. Vilkovisky, 
	Covariant perturbation theory. 2: 
	Second order in the curvature. General algorithms.
	Nucl. Phys. B {\bf 333}, 471-511 (1990).
	\href{http://dx.doi.org/10.1016/0550-3213(90)90047-H}{doi:10.1016/0550-3213(90)90047-H}
\bibitem{Migdal-book1977}
	A.B. Migdal,
	{\em Qualitative methods in quantum theory.}
	(Nauka, Moscow, 1975).
	English transl.
	(W.A. Benjamin, Reading, 1977).
\bibitem{Baltes-thesis1970}
	H.P. Baltes,
	Thermal radiation in finite cavities.
	{Helvetica Physica Acta} {\bf 45}, 481-529 (1972).
	\href{http://dx.doi.org/10.3929/ethz-a-000086038}{doi:10.3929/ethz-a-000086038}.
\bibitem{Baltes-OC1970}
	H.P. Baltes and F.K. Kneub\"uhl,
	Spectral density, thermodynamics and temporal coherence 
	of non-Planckian black body radiation for small cavities.
	{Optics Comm.} {\bf 2}, 14-16 (1970).
	\href{http://dx.doi.org/10.1016/0030-4018(70)90019-2}{doi:10.1016/0030-4018(70)90019-2}
\bibitem{Baltes-PLA1969}
	H.P. Baltes and F.K. Kneub\"uhl,
	Spectral distribution of cavity black body radiation in the far infra red.
	Phys. Lett. A {\bf 30}, 360-362 (1969).
	\href{http://dx.doi.Org/10.1016/0375-9601(69)90850-0}{doi:10.1016/0375-9601(69)90850-0}
\bibitem{Baltes-OC1971}
	H.P. Baltes and F.K. Kneub\"uhl,
	Surface area dependent corrections in the theory of black body radiation.
	Optics Comm. {\bf 4}, 9-12 (1971).
	\href{http://dx.doi.org/10.1016/0030-4018(71)90116-7}{doi:10.1016/0030-4018(71)90116-7}
\bibitem{Baltes-AP1973}
	H.P. Baltes,
	Deviations from the Stefan Boltzmann law at low temperatures.
	{Applied Phys.} {\bf 1}, 39-43 (1973).
	\href{http://dx.doi.org/10.1007/BF00886803}{doi:10.1007/BF00886803}
\bibitem{Baltes-PRA1972}
	H.P. Baltes,
	Asymptotic eigenvalue distribution for the wave equation 
	in a cylinder of arbitrary cross section.
	{Phys. Rev. A} {\bf 6}, 2252-2257 (1972).	
	\href{http://dx.doi.org/10.1103/PhysRevA.6.2252}{doi:10.1103/PhysRevA.6.2252}
\bibitem{Case-PRA1970}
	K.M. Case and S.C. Chiu,
	Electromagnetic fluctuations in a cavity.
	{Phys. Rev. A} {\bf 1}, 1170-1174 (1970).
	\href{http://dx.doi.org/10.1103/PhysRevA.1.1170}{doi:10.1103/PhysRevA.1.1170}
\bibitem{Balian-AP1971}
	R. Balian and C. Bloch,
	Distribution of eigenfrequencies for the wave equation in a finite domain. 	II.
	Electromagnetic field. Riemannian spaces.
	Ann. Phys. (N.Y.) {\bf 64}, 271-307 (1971).
    \href{http://dx.doi.org/10.1016/0003-4916(71)90286-7}{doi:10.1016/0003-4916(71)90286-7}
\bibitem{Kac-AMM1966}
	M. Kac,
	Can one hear the shape of a drum?
	Amer. Math. Month. {\bf 73}, No. 4, Part 2, 1-23 (1966).
	\href{http://www.jstor.org/stable/2313748}{www.jstor.org/stable/2313748}    
\bibitem{Quinn-PTRSLA1985}
	T.J. Quinn and J.E. Martin,
	A radiometric determination of the Stefan-Boltzmann constant 
	and thermodynamic temperatures between  
	$- 40^{\circ}C$ and $+100^{\circ}C$.
	{Phil. Trans. Roy. Soc. Lond. A}, {\bf 316}, 85-189 (1985).
	\href{http://dx.doi.org/10.1098/rsta.1985.0058}{doi:10.1098/rsta.1985.0058}
\bibitem{Gillham-PRSA1962}
	E.J. Gillham,
	Recent investigations in absolute radiometry.
	Proc. R. Soc. Lond. A {\bf 269}, 249-276 (1962).
	\href{http://dx.doi.org/10.1098/rspa.1962.0174}{10.1098/rspa.1962.0174}
\bibitem{Guild-PRSA1937}
	J. Guild,
	Investigations in absolute radiometry.
	Proc. R. Soc. Lond. A {\bf 161}, 1-38 (1937).
	\href{http://dx.doi.org/10.1098/rspa.1937.0129}{doi:10.1098/rspa.1937.0129}
\bibitem{Blevin-Metro1971}
	W.R. Blevin and W.J. Brown,
	A precise measurement of the Stefan-Boltzmann constant.
	Metrologia {\bf 7}, 15-29 (1971).
	\href{http://dx.doi.org/0026-1394/7/1/003}{doi:0026-1394/7/1/003}
\bibitem{Pathria-PRA1973}
	A.N. Chaba and R.K. Pathria,
	Edge and curvature effects in Weyl's problems.
	Phys. Rev. A {\bf 8}, 3264-3265 (1973).
	\href{http://dx.doi.org/10.1103/PhysRevA.8.3264}{doi:10.1103/PhysRevA.8.3264}
\bibitem{Martin-Metro1988}
	J.E. Martin, T.J. Quinn, B. Chu,
	Further measurements of thermodynamic temperature using 
	a total radiation thermometer: the range $-130^{\circ}C$  to $+60^{\circ}C$.
	Metrologia {\bf 25}, 107-112 (1988).
	\href{http://dx.doi.org/10.1088/0026-1394/25/2/008}{doi:10.1088/0026-1394/25/2/008}
\bibitem{Ward-Metro1979}
	S.D. Ward and  J.P. Compton,
	Intercomparison of platinum resistance thermometers and T68 calibrations.
	Metrologia {\bf 15}, 31-46 (1979).
	\href{http://dx.doi.org/0026-1394/15/1/003}{doi:0026-1394/15/1/003}
\bibitem{Preston-Metro1990}
	H. Preston-Thomas,
	The International Temperature Scale of 1990 (ITS-90).
	Metrologia {\bf 27}, 3-10 (1990).
	\href{http://dx.doi.org/0026-1394/27/1/002}{doi:0026-1394/27/1/002}
\bibitem{Datla-NIST1994}	
	R.U. Datla,  M.C. Croarkin, and A.C. Parr,
	Cryogenic blackbody calibrations at the National Institute of Standards Technology Low Background Infrared Calibration Facility.
	{J. Res. Natl. Inst. Stand. Technol.} {\bf 99}, 77-87 (1994);
	Available \href{http://www.nist.gov/manuscript-publication-search.cfm?pub_id=103712}{www.nist.gov}
\bibitem{Fellmuth-BIPM2012}
	B. Fellmuth,
	Supplementary information for the ITS-90: Chapter 1. Introduction. Edition 2012.
	Bureau International de Poids et Mesures. Available at 	\href{http://www.bipm.org/en/publications/mep_kelvin/its-90_supplementary.html}{www.bipm.org}
\bibitem{Mekhontsev-book2010}
	S.N. Mekhontsev, A.V. Prokhorov and L.M. Hanssen,
	Experimental characterization of blackbody radiation sources.
	In  Z.M. Zhang, B.K. Tsai, G. Machin (eds.), 
	{\em Radiometric Temperature Measurements: II. Applications} 
	(Elsevier \& Academic Press, San Diego, 2010), pp. 57-136.  
	\href{http://dx.doi.org/10.1016/S1079-4042(09)04302-1}{doi:10.1016/S1079-4042(09)04302-1}
\bibitem{BIPM-2008}
	B. Khlevnoy, 
	CCPR S1 supplementary comparison spectral radiance (220 to 2500 nm).
	Metrologia  {\bf 45}, 02001 (2008).
	\href{doi:10.1088/0026-1394/45/1A/02001}{doi:10.1088/0026-1394/45/1A/02001}.
	Bureau International de Poids et Mesures:
	\href{http://www.bipm.org/utils/common/pdf/final_reports/PR/S1/CCPR-S1.pdf}{CCPR-S1.pdf}
\bibitem{Quinn-book1990}
	T.J. Quinn,
	{\em Temperature}, second ed.
	(Academic Press, San Diego, 1990), sect. 7.6.
\bibitem{NIST-CoBB}
	R.D. Lee,
	Construction and operation of a simple high-precision 
	copper-point blackbody and furnace.
	NBS Technical Note 483. (National Bureau of Standards, Washington DC, 1969). 
	\href{http://archive.org/details/constructionoper483leer}{archive.org/details/constructionoper483leer}
\bibitem{Polaron}
	Polaron Special Lamps Division, 
	{\em Secondary standard lamps for optical radiation,
	 photometric and pyrometric measurements} (2013). Available at
	\href{http://www.coopercontrols.co.uk/components/spec_lamps.htm}{www.coopercontrols.co.uk}
\bibitem{Han-OC2013}
	F. Han, X. Sun, L. Wu, and Q. Li,
	Dual-wavelength orthogonally polarized radiation generated by a tungsten thermal source.
	Optics Express  {\bf 21}, 28570-28582 (2013).
	\href{http://dx.doi.org/10.1364/OE.21.028570}{doi:10.1364/OE.21.028570}
\bibitem{Sounders-IJT2011}
	P. Saunders,
	Correcting radiation thermometry measurements for the Size-of-Source effect.
	Int. J. Thermophysics {\bf 32}, 1633-1654 (2011).
    \href{http://dx.doi.org/10.1007/s10765-011-0988-9}{doi:10.1007/s10765-011-0988-9}
\bibitem{Bloembergen-Metro2009}
	P. Bloembergen,
	Analytical representations of the size-of-source effect.
	Metrologia {\bf 46}, 534-543 (2009).
	\href{http://dx.doi.Org/10.1088/0026-1394/46/5/018}{doi:10.1088/0026-1394/46/5/018}
\bibitem{Bimonte-NJP2009}   
	G. Bimonte, L. Cappellin, G. Carugno, G. Ruoso, and D. Saadeh,
	Polarized thermal emission by thin metal wires.
	New J. Phys. {\bf 11}, 033014(11) (2009).
	\href{http://dx.doi.Org/10.1088/1367-2630/11/3/033014}{doi:10.1088/1367-2630/11/3/033014}
\bibitem{Rytov-book1978}
 	S.M. Rytov, Y.A. Kravtsov, and V.I. Tatarskii, 
  	{\em Introduction to statistical radiophysics. 2. Random fields}, second ed.
 	(Nauka, Moscow, 1976), in Russian.
 	S.M. Rytov, Y.A. Kravtsov, and V.I. Tatarskii,
  	{\em Principles of statistical radiophysics. V. 3. Elements of random fields}, second ed.
 	(Springer-Verlag, Berlin, 1987).
 \bibitem{Kardar-PRE2012}
	V.A. Golyk, M.Kr\"uger, and M. Kardar,
	Heat radiation from long cylindrical objects.
	Phys. Rev. E {\bf 85}, 046603 (2012).
	\href{http://dx.doi.org/10.1103/PhysRevE.85.046603}{doi:10.1103/PhysRevE.85.046603}
\bibitem{Parr-NIST2007}
	C.E. Gibson, G.T. Fraser, A.C. Parr, H.W. Yoon,
    Once is enough in radiometric calibrations.
    J. Res. Nat. Inst. Stand. Tech. {\bf 112}, 39-51 (2007);
    Available \href{http://www.nist.gov/manuscript-publication-search.cfm?pub_id=841160}{www.nist.gov}
\bibitem{Maslov-RJMP2007}
	V.P. Maslov, 
	Revision of probability theory from the point of view of quantum statistics.
	Russ. J. Math. Phys. {\bf 14}, 66-95 (2007).
	\href{http://dx.doi.org/10.1134/S106192080701005}{doi:10.1134/S106192080701005}
\bibitem{Erdosh-BAMS1946}
	P. Erd\"osh,
	On asymptotic formulas in the theory of partitions.
	Bull. Amer. Math. Soc. {\bf 52}, 185-188 (1946).
	\href{http://dx.doi.org/10.1090/S0002-9904-1946-08540-7}
	{doi:10.1090/S0002-9904-1946-08540-7}
\bibitem{Pathria-book1996}
	H.K. Pathria,
	{\em Statistical mechanics}, second ed.
	(Butterworth-Heinemann, Oxford, 1997).
\bibitem{Landsberg-JPA1989}
	P.T Landsberg and A. De Vos,
	The Stefan-Boltzmann constant in {\em n}-dimensional space.
	J. Phys. A: Math. Gen. {\bf 22} 1073-1084 (1989).
	\href{http://iopscience.iop.org/0305-4470/22/8/021}{iopscience.iop.org/0305-4470/22/8/021}
\bibitem{Stefan-1879}
	J. Stefan,
	\"Uber die Beziehung zwischen der W\"armestrahlung und der Temperatur
	[About the relationship between the thermal radiation and temperature].
	Wien. Berichte {\bf LXXIX}, 391-429 (1879).
	Abstract at \href{http://www.emis.de/cgi-bin/jfmen/MATH/JFM/quick.html?first=1&maxdocs=13&type=html&an=JFM\%2011.0790.01&format=complete}{www.emis.de}
\bibitem{Crepeau-ETFS2007}
	J. Crepeau,
	Josef Stefan: His life and legacy in the thermal sciences.
	Exp. Thermal Fluid Science {\bf 31}, 795-803 (2007).
	\href{http://dx.doi.org/10.1016/j.expthermflusci.2006.08.005}{doi:10.1016/j.expthermflusci.2006.08.005}
\bibitem{Dougal-PhysEd1979}
	R.C. Dougal,
	The centenary of the fourth-power law.
	Phys. Educ. {\bf 14}, 234-238 (1979).
	\href{http://dx.doi.org/10.1088/0031-9120/14/4/312}{doi:10.1088/0031-9120/14/4/312}
\bibitem{Becker-Metro2007}
	P. Becker, P. De Bi\'evre, K. Fujii, M. Glaeser, B. Inglis, H. Luebbig. and G. Mana,
	Considerations on future redefinitions of the kilogram, the mole and of other units.
	Metrologia {\bf 44}, 1-14 (2007).
	\href{http://dx.doi.org/10.1088/0026-1394/44/1/001}{doi:10.1088/0026-1394/44/1/001}
\bibitem{BIPM-newSI2011}
	Bureau International des Poids et Mesures (BIPM),
	On the possible future revision of the International System of Units, the SI.
	24th meeting of the General Conference on Weights and Measures, Oct. 2011.
	Available at \href{http://www.bipm.org/utils/en/pdf/24_CGPM_Resolution_1.pdf}{www.bipm.org}	
\bibitem{Planck-AdP1900}
	M. Planck,
	\"Uber irreversible Strahlungsvorg\"ange
	[About irreversible radiation processes].
	Annalen der Physik {\bf 306} (1), 69-122 (1900).
	\href{http://dx.doi.org/10.1002/andp.19003060105}{doi:10.1002/andp.19003060105}
\bibitem{Shukla-book2002}
	P.K. Shukla and A.A. Mamun,
	{\em Introduction to dusty plasma physics}
	(IOP Publishing, Bristol, 2002)
\bibitem{Bacharis-PPlasmas2010}
	M. Bacharis, M. Coppins, and J.E. Allen,
	Dust in tokamaks: An overview of the physical model of the dust in tokamaks code.
	Phys. Plasmas {\bf 17}, 042505(11) (2010).
	\href{http://dx.doi.org/10.1063/1.3383050}{doi:10.1063/1.3383050}
\end{thebibliography}
\end{document}